\documentclass[sigconf,screen]{acmart}

\usepackage{subcaption}
\usepackage{enumitem}
\AtBeginDocument{%
  \providecommand\BibTeX{{%
    \normalfont B\kern-0.5em{\scshape i\kern-0.25em b}\kern-0.8em\TeX}}}

\setcopyright{acmcopyright}
\copyrightyear{2023}
\acmYear{2023}
\acmDOI{XXXXXXX.XXXXXXX}

\acmConference[CF '23]{ACM International Conference on Computing Frontiers }{May 09--11,
  2023}{Bologna, IT}
\acmPrice{15.00}
\acmISBN{978-1-4503-XXXX-X/18/06}

\begin{document}

\title{Fault Awareness in the MPI 4.0 Session Model}

\author{Roberto Rocco}
\orcid{0000-0002-0223-2900}
\affiliation{%
  \institution{Dipartimento di Elettronica, Informazione e Bionigegneria Politecnico di Milano}
  \streetaddress{Via Giuseppe Ponzio, 34}
  \city{Milan}
  \country{Italy}
}
\email{roberto.rocco@polimi.it}

\author{Gianluca Palermo}
\affiliation{%
  \institution{Dipartimento di Elettronica, Informazione e Bionigegneria Politecnico di Milano}
  \streetaddress{Via Giuseppe Ponzio, 34}
  \city{Milan}
  \country{Italy}
}
\email{gianluca.palermo@polimi.it}

\author{Daniele Gregori}
\affiliation{
  \institution{E4 Computer Engineering Spa}
  \streetaddress{Viale Martiri della Libertà, 66}
  \city{Scandiano (RE)}
  \country{Italy}
}
\email{daniele.gregori@e4company.com}


\begin{abstract}
The latest version of MPI introduces new functionalities like the Session model, but it still lacks fault management mechanisms. Past efforts produced tools and MPI standard extensions to manage fault presence, including ULFM. These measures are effective against faults but do not fully support the new additions to the standard.
In this paper, we combine the fault management possibilities of ULFM with the new Session model functionality introduced in version 4.0 of the standard. We focus on the communicator creation procedure, highlighting criticalities and proposing a method to circumvent them. The experimental campaign shows that the proposed solution does not significantly affect applications' execution time and scalability while better managing the insurgence of faults.
\end{abstract}




\maketitle

\section{Introduction}\label{sec:intro}

The increase in computation capabilities is leading the innovation in the High-Performance Computing (HPC) field. The future (and present) exascale clusters introduce new challenges for application developers, system management and framework designers, making even consolidated tools for the HPC field evolve. One of these tools is the Message Passing Interface (MPI), the de-facto standard for inter-process communication. With its latest version (v4.0) \cite{mpi40}, the MPI standard introduces functionalities that impact how the application approaches communication. The new Session model \cite{holmes2016mpi} enables communication initialisation in multiple independent instances, achieving decoupling across different application modules. It also avoids the creation of a global communicator when unneeded, which can bring savings in terms of synchronisation and initialisation overhead.

The higher number of nodes involved in the execution also makes fault occurrence more frequent. The MPI standard, however, does not specify the application behaviour after a fault manifests. Many efforts proposed tools to introduce fault management techniques inside MPI applications, with the User Level Fault Mitigation (ULFM) extension \cite{bland2013post} the one currently receiving the most attention. It proposes a set of additional functionalities to repair the communicators involved with the fault. The ULFM extension targets the classical initialisation model, but its approach can also benefit applications using the Session model.

In this work, we further analyse ULFM usage in applications employing the Session model. We show the limitations and criticalities of a naive approach and propose the Horizon set usage, which maintains the benefits of the Sessions model while achieving fault support. The experimental campaign proved that our solution does not compromise the scalability and performance of the application while correctly managing faults.

The paper is structured as follows: Section~\ref{sec:background} explains the fundamental concepts of the Session model and ULFM extension alongside the previous work; Section~\ref{sec:init} explores the fault impact on the initialisation phase for both models; Section~\ref{sec:Horizon} discusses our approach to the problem; and Section~\ref{sec:experimental} shows the experimental campaign that validates our solution. Lastly, Section~\ref{sec:conclusion} concludes the paper.

\section{Background}\label{sec:background}

In this section, we discuss the main concepts and works that influence this effort. We cover the MPI Session model basics in Section~\ref{sec:background:basics}, the fundamental ULFM principles in Section~\ref{sec:background:ulfm}, and the previous work in Section~\ref{sec:background:related}.

\subsection{The MPI Session model}\label{sec:background:basics}

The main driver behind the Session model introduction is the communication isolation between different application modules. The problem with the classical MPI approach (the World model) resides in the uniqueness of the initialisation call. 
If many application components, modules or threads want to use MPI functionalities, they have to coordinate such that only one calls the \texttt{MPI\_Init} function, compromising the isolation between the parts. On the other hand, the Session model allows each component, module or thread to initialise and manage its communication independently from the others. This possibility allows for modularity inside the software and reduces the coordination need, making the Session approach more scalable and better fitting the exascale scenario. 

\begin{figure}
    \centering
    \includegraphics[width=0.6\linewidth]{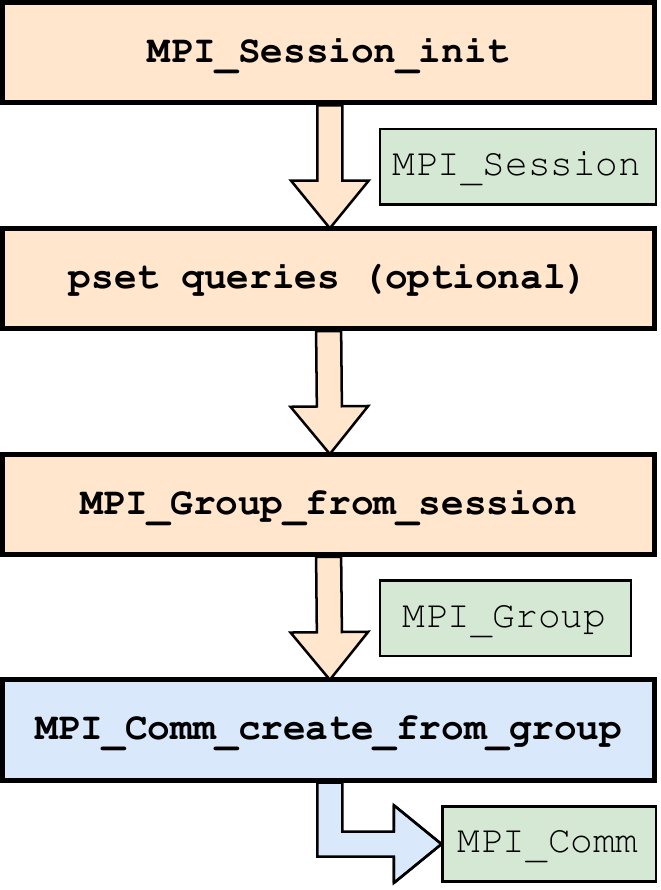}
    \caption{The Session model execution flow, from the Session creation to the communicator generation. Boxes in orange represent local operations, the ones in green the produced handles, and the ones in blue functions that require synchronisation.}
    \label{fig:session}
\end{figure}

While the World model initialisation process happens entirely through the \texttt{MPI\_Init} call, the Session model requires a more complex procedure. Figure~\ref{fig:session} summarises the Session model creation flow. Most of the functions are local and thus require no communication, except for the \texttt{MPI\_Comm\_create\_from\_group} call, which needs participation from all the processes part of the future communicator. After the communicator generation, the application operates analogously to the World model.

\subsection{The User Level Fault Mitigation extension}\label{sec:background:ulfm}

The MPI standard does not specify the execution behaviour after the incurrence of a fault, practically precluding any fault management technique. Many efforts proposed solutions to circumvent this limitation, with ULFM \cite{bland2013post} being the one currently receiving the most attention. It is an MPI standard extension that allows the execution to continue even if a process stops working and introduces calls to propagate faults, remove them and reach an agreement even in a faulty scenario (with the \textit{revoke}, \textit{shrink} and \textit{agree} functions respectively). However, faults can impact the execution's correctness, requiring additional recovery techniques. 

ULFM considers faults as an abrupt and unplanned termination of computation. Other fault types, like silent data corruption and timing errors, usually require ad-hoc measures and are thus excluded from the ULFM analysis. Moreover, ULFM does not provide any recovery mechanism, expecting application developers to choose it depending on their needs. 

The ULFM fault management features apply from communicator creation to its destruction. This assumption makes the ULFM approach viable in the Session model since the behaviour after communicator creation is the same as the World one. However, ULFM does not specify the application behaviour if faults happen before and during the \texttt{MPI\_Init} call. This limitation does not heavily impact the World model since no MPI routine is callable before initialisation, so applications perform it as soon as possible. On the other hand, due to the added flexibility from the Session model, we can avoid calling the \texttt{MPI\_Init} function or invoke the Session initialisation procedure multiple times and further in the execution, extending the period for a possible undefined behaviour. In this work, we try to enlarge the ULFM-defined behaviour, analysing the Session model initialisation phase.

\subsection{Related work}\label{sec:background:related}

Aside from the ULFM extension, the other main direction in the MPI fault tolerance field is the Reinit proposal \cite{laguna2014evaluating,chakraborty2020ereinit}. Its main feature is the possibility to perform the \texttt{MPI\_Init} function multiple times. In case of failure, no repair process is necessary since re-initialising the communication is sufficient. 
The approach provided by the Reinit solution is similar to the multiplicity of the sessions but focuses only on the World model and would bring no additional benefit in the Session one compared to ULFM.

Out of all the efforts leveraging the ULFM extension, \cite{rocco2022fault} analysed the \texttt{MPI\_Comm\_create\_from\_group} function, proposing a solution that enables the completion of the call even with faults among the involved processes. They introduce a Liveness Discovery Algorithm that queries the execution and proactively discovers failures. This solution is promising since it deals with a critical function of the Session initialisation process but limits its scope to the World model.

A completely different approach comes from the Session Working Group (responsible for the Session model development inside the MPI standard). They are currently working on extending the idea behind the Session model to obtain isolated communication \textit{Bubbles} \cite{huber2022towards}. This work should open up for malleability properties in MPI, allowing dynamic reconfiguration of the execution. Moreover, those benefits also affect fault tolerance since they allow handling faults as resource changes. While being a promising direction, the fault tolerance analysis requires additional efforts: the Bubble idea can represent the faults' consequences but needs effective detection and propagation mechanisms. This work is orthogonal to the Bubbles proposal since it operates at different levels while keeping the approaches compatible.

\section{Failures in the initialisation flows}\label{sec:init}

In this section, we analyse the effects of faults on the initialisation flows of the two MPI execution models. For reference, we use an MPI implementation with ULFM features enabled. Otherwise, fault incurrence would have just stopped the execution regardless of our measures. 

In the World model, the initialisation procedure consists only of the \texttt{MPI\_Init} function. It involves all the processes part of the execution and produces the starting communicators. If someone does not participate in the call, the others will wait until it joins. If that process cannot join, the execution enters a deadlock state, which is also the case if some process fails before the function invocation. From the execution viewpoint, it is impossible to decide whether the function will eventually terminate or is in a deadlock state: this decision is analogous to the \textit{Halting Problem} \cite{turing1936computable}. The deadlock-vulnerable function is, however, the first call of every MPI application, minimising the vulnerability time window. Overall, the deadlock risk is acceptable, so ULFM can focus on handling faults in the rest of the application.

In the Session model, we face a different situation: almost all the functions of the initialisation flow are local, thus requiring no communication and providing a result even with faults. The only non-local operation is the \texttt{MPI\_Comm\_create\_from\_group} call, which needs the participation of all the processes involved in the communication creation. Similarly to the \texttt{MPI\_Init} function, the communicator creation call cannot distinguish between faults and delays and exposes a deadlock whenever a process does not participate. Differently from the \texttt{MPI\_Init} function, however, the initialisation is not unique and can happen at any point of the application execution, even multiple times per session. These considerations imply that the deadlock risk has a high impact and moderate probability of verifying, making the problem relevant.

\section{The Horizon sets} \label{sec:Horizon}

The effort \cite{rocco2022fault} analysed the \texttt{MPI\_Comm\_create\_from\_group} function from the fault management perspective, removing the deadlock eventuality from its behaviour. The authors leveraged a Liveness Discovery Algorithm (LDA) to explore the group and remove failed processes. Their approach requires an already-created communicator, including all the processes that intend to be part of the new one, so they opted for \texttt{MPI\_COMM\_WORLD}. Since \texttt{MPI\_COMM\_WORLD} is usable only in the World model, that approach is valid only if another communicator is available. However, to create such a communicator, we can only use the \texttt{MPI\_Comm\_create\_from\_group} function, which can still deadlock. The situation resembles a \textit{chicken-egg} problem. Nonetheless, if we provide a correct communicator, we can use the LDA algorithm and remove the deadlock eventuality. This analysis aims to create such a communicator, thus reducing the deadlock vulnerability but preserving the application performance.

To better visualise the problem, we introduce a set of notions. For a given group of processes $G$, we define as $Create(C_G)$ the call which produces a communicator $C_G$ with all the processes part of $G$. Moreover, we refer to the \textit{superset} $S(G)$ as the set of communicators containing at least all the processes of $G$. From these definitions, we can deduce these simple assumptions:
\begin{enumerate}[label=\textbf{A.\arabic*}]
\item \label{Horizon:A1} $ \forall G : C_G \in S(G) \land \texttt{MPI\_COMM\_WORLD} \in S(G)$
\item \label{Horizon:A2} $ \forall G1,G2 : C_{G2} \in S(G1) \Rightarrow G2 \supseteq G1$
\end{enumerate}
The first assumption guarantees that the superset of each group is never empty, while the other defines the relationship between different supersets.
\begin{figure}[t]
    \centering
    \begin{subfigure}[b]{\linewidth}
        \centering
        \resizebox{0.7\linewidth}{!}{
            \includegraphics[width=\linewidth]{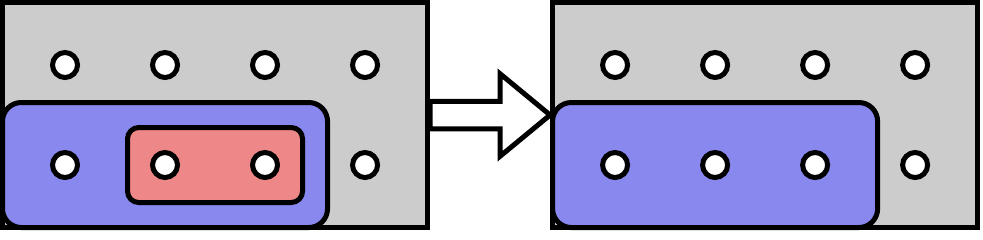}
        }
        \caption{new communicator included in Horizon}
    \label{fig:Horizon:a}
    \vspace{5pt}
    \end{subfigure}
    \vspace{5pt}
    \begin{subfigure}[b]{\linewidth}
        \centering
        \resizebox{0.7\linewidth}{!}{
            \includegraphics[width=\linewidth]{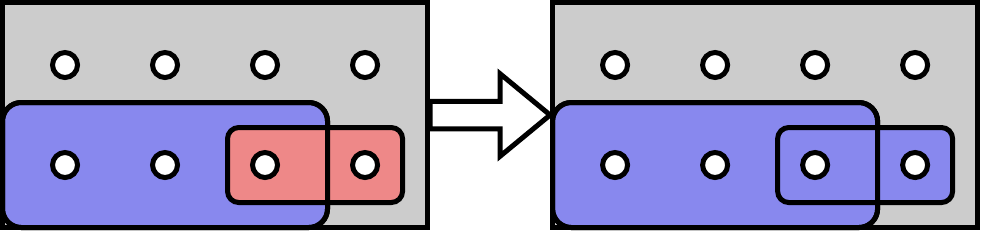}
        }
        \caption{new communicator intersects Horizon}
    \label{fig:Horizon:b}
    \end{subfigure}
    \begin{subfigure}[b]{\linewidth}
        \centering
        \resizebox{0.7\linewidth}{!}{
            \includegraphics[width=\linewidth]{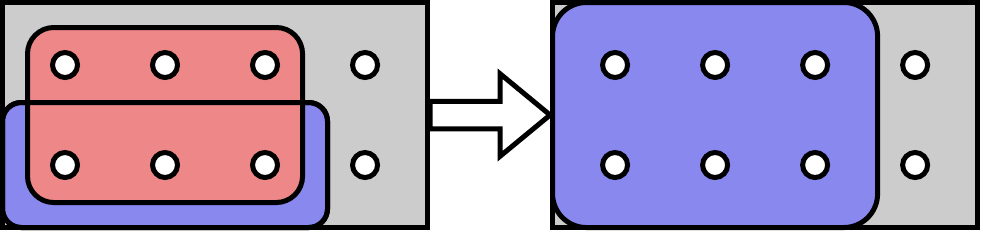}
        }
        \caption{Horizon included by the new communicator}
    \label{fig:Horizon:c}
    \end{subfigure}
    \caption{Evolution of the Horizon set (in blue) when introducing a new communicator (in red). The grey rectangle represents the global communicator, each white circle a process inside it. The behaviour depends on whether there is an inclusion relationship (cases~\ref{fig:Horizon:a} and~\ref{fig:Horizon:c}) or not (case~\ref{fig:Horizon:b}).}
    \label{fig:Horizon}
\end{figure}
The requirement for the Liveness Discovery Algorithm (LDA) and the deadlock removal is the existence of a call to create a communicator part of the superset of the group passed as a parameter. Formally:
\begin{multline}
LDA\,applicable\,for\,Create(C_G) \Leftrightarrow \\
\exists K \in S(G)\,|\,Create(K)\,prior\,to\,Create(C_G)
\label{eq:problem}
\end{multline}
Given the above analysis, if we want to minimise the deadlock eventuality, we must define a communicator set $W$ such that $\forall G\,\exists K \in W\,|\,K \in S(G)$. We should initialise all the communicators inside $W$ before the first communicator creation call so that, for each successive $Create(C_G)$, Equation~\ref{eq:problem} is valid. Moreover, following the Assumption~\ref{Horizon:A1}, we can deduce that the minimal $W$ set contains only a communicator equivalent to \texttt{MPI\_COMM\_WORLD}. Given this conclusion, then \textit{naive} solution to the deadlock vulnerability problem is the generation of a global communicator during the first call of \texttt{MPI\_Session\_init}, which will happen before any other operation. This solution, however, features many problems:
\begin{itemize}
\item The function \texttt{MPI\_Session\_init} is local, thus requiring no communication between processes. Communicator creation needs coordination, changing the function's characteristics.
\item The \texttt{MPI\_Session\_init} invocation must happen as soon as possible to minimise the vulnerability window, compromising the Session model flexibility.
\item One of the Session model benefits consists in its scalability due to not requiring the application to initialise the entire global communicator, but just the needed portions. This solution loses that benefit.
\end{itemize}
This \textit{naive} solution is thus insufficient, but it is a good starting point. Its main weakness is its need to create a global communicator, regardless of the characteristics of the application. Moreover, its use costs some of the Session model benefits, so we opted for a different approach.

While global communicator usage is the safest option, it is not the only one. To further the analysis, we propose the use of \textit{Horizon set}s $H(t)$, defined such that the following formulation is valid:
\begin{multline}
\forall G : Create(C_G)\,prior\,to\,t \Rightarrow \\
\exists C \in H(t)\,|\,C \in S(G) \land Create(C)\,prior\,to\,t
\label{eq:definition}
\end{multline}
In other words, at each time $t$, a Horizon set contains at least an element of the superset of all the groups passed as a parameter in a previous communicator creation call. Moreover, combining the Equation~\ref{eq:definition} with the Equation~\ref{eq:problem}, we deduce that the LDA is applicable if there exists an element of the superset of the group passed as a parameter inside a Horizon set. Formally:
\begin{multline}
\forall Create(C_G) : Create(C_G)\,happens\,at\,t+1 \\
LDA\,applicable\,for\,Create(C_G) \Leftrightarrow \\
\exists K \in H(t)\,|\,K \in S(G)
\label{eq:thesis}
\end{multline}
The Equation~\ref{eq:thesis} links Horizon sets to our problem, so we can reason on the formers to solve the latter.

Horizon sets evolve with time, growing in size with new communicators. The collection of all the communicators created before time t is a Horizon set, and we can use it to decide whether the LDA algorithm is applicable. However, that Horizon set may not be the lowest cardinality one. We define the minimal Horizon set as $H^{*}(t)$. Focusing on the minimal Horizon set is mandatory if we want to reduce the overhead due to communicator management.

We can analyse the evolution of $H^{*}(t)$ by considering all the possible topology cases of communicator creations, as shown in Figure~\ref{fig:Horizon}. In general, the addition of a communicator to the minimal Horizon set must not take place if another communicator containing it already exists (Figure~\ref{fig:Horizon:a}). On the other hand, the newly inserted communicator removes all the communicators it includes from the Horizon set (Figure~\ref{fig:Horizon:c}). Following these concepts, we can keep track of $H^{*}(t)$ and use it to remove the deadlock eventuality from some communication creation calls.

Inside the application code, calls to $Create(C_G)$ refer to invocations of functions \texttt{MPI\_Comm\_create\_from\_group} or \texttt{MPI\_Init} (which creates \texttt{MPI\_COMM\_WORLD}). By keeping track of those functions, we can update the minimal Horizon set correctly and use it to limit the deadlock vulnerability of the Session model initialisation flow. Moreover, we provide a new ad-hoc collective function \texttt{MPIX\_Horizon\_from\_group}, which includes the communicator of the group passed as a parameter inside the Horizon set. While its behaviour is analogous to the communicator creation functions, it allows the user to provide the application communication intent.


\section{Experimental Results} \label{sec:experimental}

We integrated the proposed Horizon set management functionalities inside the Legio library \cite{rocco2021legio, rocco2022fault} since it already implements the LDA algorithm and operates at the same abstraction level. All the added functionalities leverage the PMPI layer to avoid tampering with the application code directly.

The experimental campaign evaluates the proposed solution overhead and scalability in a real-world scenario. In particular, we integrate the solution with the embarrassingly parallel (EP) and data traffic (DT) benchmarks part of the NAS collection \cite{bailey1995parallel} and execute them on the IT4Innovations Karolina cluster, featuring nodes with 2 x AMD Zen 2 EPYC™ 7H12, 2.6 GHz processors and 256 GB of RAM, each running 128 MPI processes. We use the latest version of OpenMPI featuring ULFM (v5.0.0) and implementing MPI standard 4.0. We changed the two benchmarks to leverage the Session model. We use a "C" size workload for both applications, the maximum available for their executions.

\begin{figure}[t]
    \centering
    \begin{subfigure}[b]{0.47\linewidth}
        \centering
            \includegraphics[width=\linewidth]{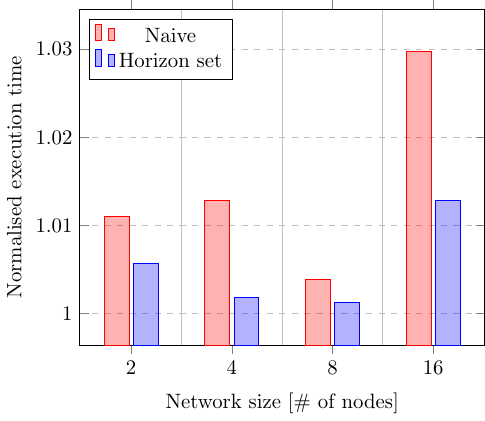}
        \caption{DT benchmark}
    \label{fig:oh_data:dt}
    \end{subfigure}
    \hspace{5pt}
    \begin{subfigure}[b]{0.47\linewidth}
        \centering
            \includegraphics[width=\linewidth]{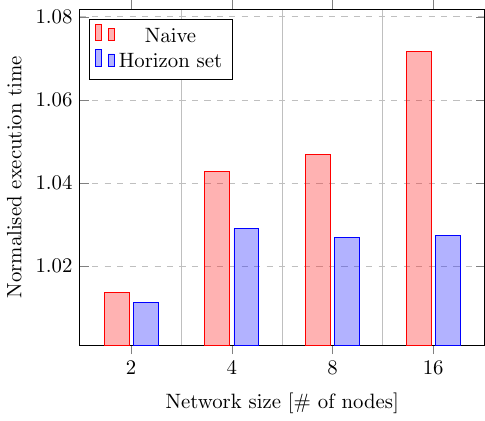}
        \caption{EP benchmark}
    \label{fig:oh_data:ep}
    \end{subfigure}
    \caption{Normalised execution times of the two benchmarks over different network sizes.}
    \label{fig:oh_data}
\end{figure}

We ran tests using different network sizes to evaluate the scalability impact of our solution. We ran both applications with the naive solution, with Horizon set support, and without fault management support. For each configuration, we measured the total time to complete ten executions. Figure~\ref{fig:oh_data} shows the benchmarks' execution time, normalised to the values without fault management support. 

The results from the two benchmarks' executions differ due to their peculiarities. In particular, we used the EP benchmark (Figure~\ref{fig:oh_data:ep}) to stress the solutions' scalability, while DT focuses on the overall execution impact (Figure~\ref{fig:oh_data:dt}). We can see from the figures that the naive approach always has more overhead than the proposed Horizon set solution. Moreover, the naive approach overhead grows with the network size, as visible in Figure~\ref{fig:oh_data:ep}. On the other hand, the Horizon set solution does not heavily impact the application scalability since it does not require the initialisation of the global communicator present in the naive approach.

Aside from performance-oriented experiments, we did functional tests to prove the resiliency of our approach. Those tests showed that an application leveraging our solutions is more deadlock-resilient, managing to complete even with faults in the execution.

\section{Conclusion} \label{sec:conclusion}

In this effort, we analysed and limited the impact of faults in MPI applications using the Session model. We motivated the need for Horizon sets, and we analysed their behaviour. We showed our proposed implementation of the Horizon set management functionalities and tested it using benchmarks in a real-world scenario. Results showed that the Horizon set solution does not compromise the performance and scalability of executions, reduces the deadlock vulnerability to faults and performs better than the naive one.

Future work in this area includes interfacing with the MPI Bubbles proposal and additional evaluation on new benchmarks that could further stress the capabilities of the Session model. 

\bibliographystyle{ACM-Reference-Format}
\bibliography{paper}

\end{document}